# Dual-Class Stocks: Can They Serve as Effective Predictors?

Veli SAFAK

**Abstract**
Kardemir Karabuk Iron Steel Industry Trade & Co. Inc., ranked as the 24th largest industrial company in Turkey, offers three distinct stocks listed on the Borsa Istanbul: KRDMA, KRDMB, and KRDMD. These stocks, sharing the sole difference in voting power, have exhibited significant price divergence over an extended period. This paper conducts an in-depth analysis of the divergence patterns observed in these three stock prices from January 2001 to July 2023. Additionally, it introduces an innovative training set selection rule tailored for LSTM models, incorporating a rolling training set, and demonstrates its significant predictive superiority over the conventional use of LSTM models with large training sets. Despite their strong coherence, the study found no compelling evidence supporting the efficiency of dual-class stocks as predictors of each other's performance.

## 1. Introduction

The establishment of Kardemir, Turkey's inaugural integrated iron and steel factory, dates back to September 10, 1939, when it was initiated by İsmet İnönü, who served as Prime Minister during that time. This significant step was part of the broader national industrialization efforts championed by the republic's founder, Mustafa Kemal Atatürk. The main activity subject of the company is the production and sale of all kinds of crude iron and steel products, coke, and coke by-products. It was listed in the Borsa Istanbul on Jun 1, 1998, with 3 groups of stocks: group A (ticker: KRDMA), group B (ticker: KRDMB), and group D (ticker: KRDMD). The Group A shareholders have the right to elect 4 members to the Board of Directors, the Group B shareholders have the right to elect 2 members to the Board of Directors, and the Group D shareholders have the right to elect 1 member to the Board of Directors. Apart from this voting privilege, there are no other privileges.

This stock structure with different voting privileges is known as dual-class stock structure. There is significant cross-country evidence suggesting that investors pay a premium for stocks with voting privileges. The pioneering empirical investigation in this domain was conducted by Lease, McConnell, and Mikkelson [7], who demonstrated that higher vote shares in the United States are associated with a premium of approximately 5%. Horner [18] examined dual-class stocks in Switzerland and observes a voting premium of merely about 1%. Zingales [17] identified a substantial premium of roughly 80% in Italy. Smith and Amoako-Adu [3] detected a premium of around 19% in Canada during the period 1988-1992, which closely resembles the premium documented in Sweden by Rydqvist [14] at 15%. Additionally, Megginson [26] provided evidence of a premium of around 13% in the United Kingdom.

Voting premium is not the only interesting phenomenon about the dual class structures. There is also evidence suggesting that prices of dual-class stock also tend to exhibit high co-integration [1]. Since 2014, the relative price ratio of GOOG (without voting power) and GOOGL (with voting power) ranged between 0.9459 and 1.05. Wu [13] used the co-integration between GOOG and GOOGL and designed a pair trading strategy. Pair trading constitutes a market-neutral tactic centered on the selection of stock pairs grounded in their relative prices or alternative indicators. The primary objective is to pinpoint pairs that exhibit a substantial level of correlation or cointegration, indicative of their tendency for synchronized price movements. This strategy finds prevalent usage among hedge funds and can be further refined through the assimilation of supplementary insights, such as volatility, antipersistence, or qualitative information derived from financial reports. Diverse methodologies, spanning statistical assessments, machine learning algorithms, and genetic programming, can be employed to unearth lucrative pairs and formulate trading cues. The efficacy of pair trading extends across various asset categories and market conditions, with certain investigations intimating heightened effectiveness in periods of market decline [12], [16], [8], [9], and [5].

In this paper, I examine historical voting premium paid for Kardemir stocks. Furthermore, I analyze the relationship between their daily returns by using wavelet coherence analysis. Finally, I demonstrate the benefits of using recurrent training sets in LSTM models for financial forecasting.

## 2. Methodology

### 2.1. Data and Variables

In this study, I use daily mid-prices (in Turkish lira) for Kardemir stocks (tickers: KRDMA, KRDMB, KRDMD) between Jan 2001 and July 2023. I calculate daily mid-prices as follows:

$$HL = 0.5(high + low)$$

The daily premium of Group $i$ over Group $j$ is calculated as the percentage difference between daily mid-prices for Group $i$ and daily mid-prices for Group $j$.

$$v_{ij} = HL_i/HL_j - 1$$

### 2.2. Wavelet Coherence Analysis

In this paper, I use the continuous wavelet transform (CWT) to quantify the magnitude, direction and lead-lag effects between Kardemir stocks. This approach has a number of advantages. First, it uncovers the dynamic relationship between these stocks, allowing me to distinguish between periods at which prices are linked. Secondly, using the CWT, I can identify changes in the direction of the relationship over time. Finally, the CWT provides insights about the relationship between these stocks at different time horizons simultaneously.

According to Torrence and Campo [6], the wavelet coefficients $W_{\varepsilon,\tau}$ associated with a time series $f(t)$ are calculated as:

$$W_{\varepsilon,\tau} = \sum_{t=1}^{n} f(t) \psi^* \left[ \frac{t-\tau}{\varepsilon} \right]$$

where $*$ represent the complex conjugate, $\varepsilon > 0$ is the scale associated with the wavelet and $\tau \in [-\alpha, \alpha]$ is the window location and $1/\varepsilon$ is the normalization factor. In this study, I use Morlet wavelet with wave number $\omega_0 = 6$ following Grinsted et al. [2]. More specifically, the Morlet wavelet is formulated as:

$$\psi(t) = \pi^{0.25} e^{i\omega_0 t} e^{\frac{-t^2}{2}}.$$

The cross-wavelet power spectrum is calculated as the product of two wavelet coefficients and represents the common variation between two time series over time and scale. It is formulated as:

$$W_{\varepsilon,\tau}(f,g) = W_{\varepsilon,\tau}(f) W^*_{\varepsilon,\tau}(g).$$

Like the correlation, the wavelet squared coherency is defined by normalizing the smoothed cross-wavelet power spectrum by the smoothed wavelet power spectrum associated with the individual time series:

$$\rho^2_{\varepsilon,\tau} = \frac{\left| Q\left(\varepsilon^{-1} W_{\varepsilon,\tau}(f,g)\right) \right|^2}{\left| Q\left(\varepsilon^{-1} W_{\varepsilon,\tau}(f)\right) \right|^2 \left| Q\left(\varepsilon^{-1} W_{\varepsilon,\tau}(g)\right) \right|^2},$$

where $Q$ is the smoothing operator. By construction, $\rho^2_{\varepsilon,\tau}$ takes values between 0 and 1. It implies no comovement when $\rho^2_{\varepsilon,\tau} = 0$, and perfect comovement when $\rho^2_{\varepsilon,\tau} = 1$. To identify statistically significant squared coherency regions, I use a Monte-Carlo method with 1,000 iterations.

To uncover lead-lag effects, I use the following wavelet multi-scale phase:

$$\theta_{\varepsilon,\tau}(f,g) = tan^{-1} \left( \frac{\mathcal{I}\left(Q\left(\varepsilon^{-1} W_{\varepsilon,\tau}(f,g)\right)\right)}{\mathcal{R}\left(Q\left(\varepsilon^{-1} W_{\varepsilon,\tau}(f,g)\right)\right)} \right).$$

Here, $\mathcal{I}$ and $\mathcal{R}$ represent the imaginary and real components of the wavelet coefficients. Phase arrows are utilized within wavelet coherence plots to depict the direction of simultaneous movement and the effects of leading or lagging. Arrows pointing east (west) signify being in (out of) sync, while arrows pointing north (south) indicate that one time series leads (lags) the other. When the phase arrow points in a northeast (southeast) direction, it means that the two series are in sync, but the second one (or first one) leads the first one (or second one). Differing outcomes are conveyed by arrows facing northwest and southwest.

## 2.3. Long Short-Term Memory (LSTM)

While training a recurrent neural network, each iteration receives an update proportional to the partial derivative of the error function with respect to its current weight. When the gradient is vanishingly small, the training may slow and, in some cases, stops [23]. The long short-term memory technique [24] is developed as a potential solution for the vanishing gradient problem. The LSTM approach is widely used in predicting stock prices because of its capacity to recognize patterns and generate more accurate predictions compared to other methods [21], [11], [20], [15], and [19].

An LSTM unit consists of a cell, and within this cell, there are three gates that manage the movement of information and regulate the cell state. These gates include an input gate, an output gate, and a forget gate. The LSTM units are then interconnected, forming a chain where each individual cell acts as a memory module within the LSTM architecture. Figure 1 illustrates a standard LSTM architecture and Figure 2 shows a standard LSTM cell architecture.

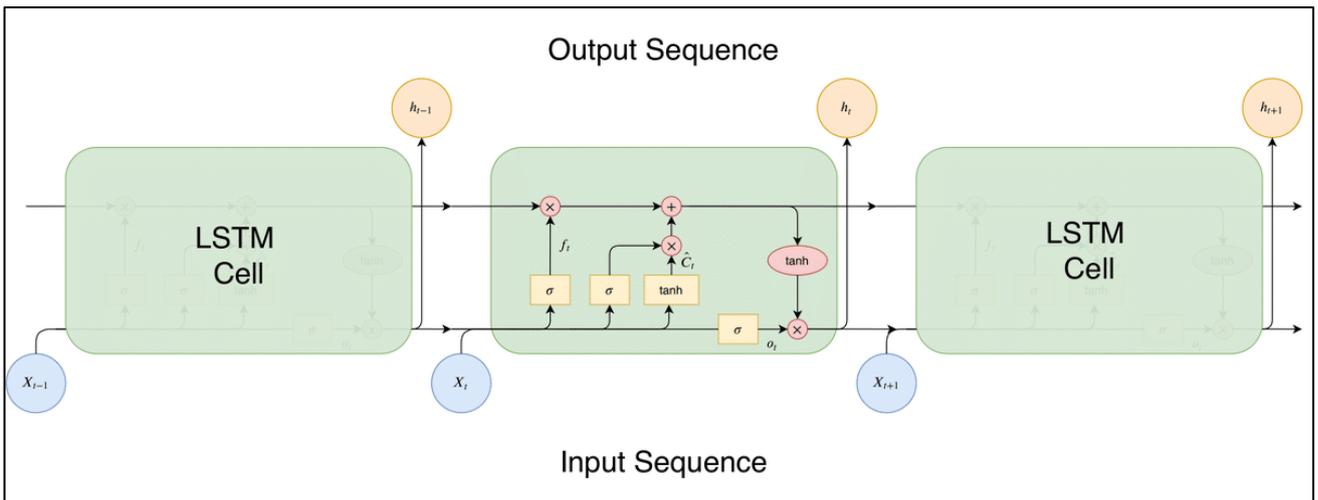

**Figure 1: LSTM Architecture (source: [25])**

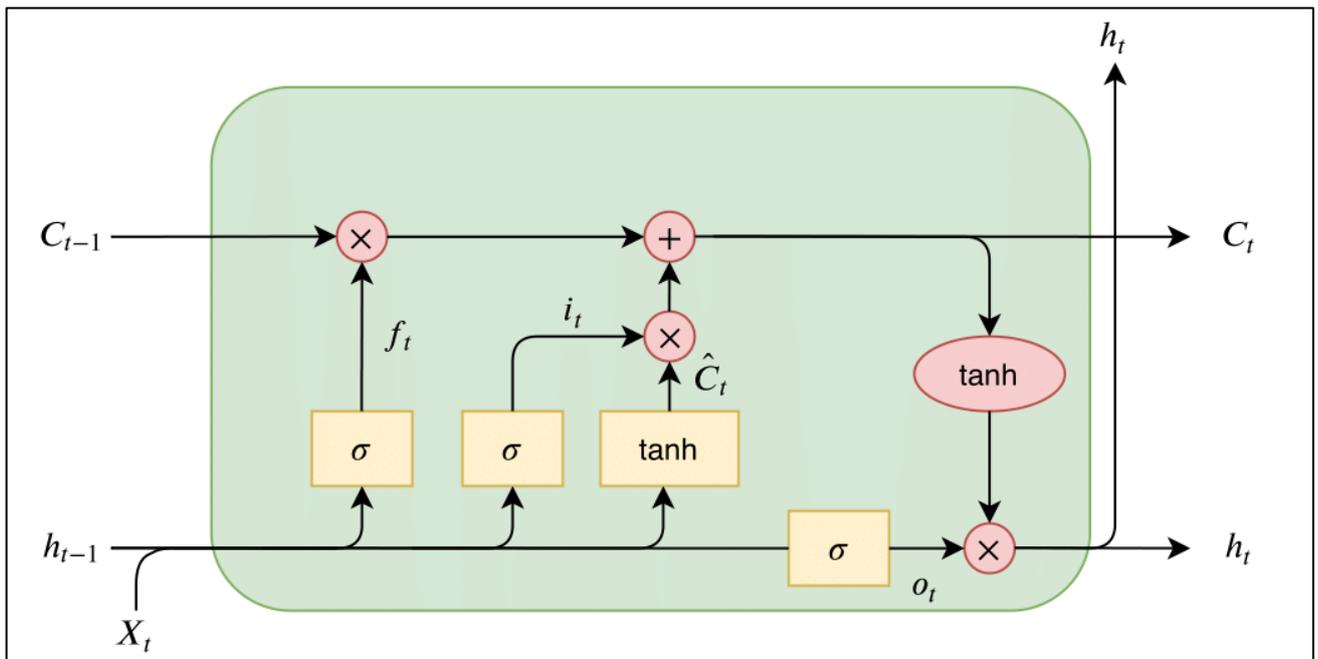

**Figure 2: LSTM Cell Architecture (source: [25])**

In Figure 2, $f_t$, $i_t$, and $o_t$ respectively represent the forget gate, input gate, and output gate. Also, $X_t$ is the input, $h_t$ is the output, $C_t$ is the cell state, and $\hat{C}_t$ is the internal cell state. Based on the input, previous output, and previous cell state ($X_t$, $h_{t-1}$, and $C_{t-1}$); $f_t$, $i_t$, $o_t$, $\hat{C}_t$, $C_t$, and $h_t$ are calculated as follow:

$$f_t = \sigma(W_f \cdot [h_{t-1}, X_t] + b_f)$$
$$i_t = \sigma(W_i \cdot [h_{t-1}, X_t] + b_i)$$
$$o_t = \sigma(W_o \cdot [h_{t-1}, X_t] + b_o)$$
$$\hat{C}_t = tanh(W_C \cdot [h_{t-1}, X_t] + b_C)$$
$$C_t = i_t \cdot \hat{C}_t + f_t \cdot C_{t-1}$$
$$h_t = o_t \times tanh(C_t)$$

Here, $\sigma$ represents the sigmoid function and $tanh$ represents the hyperbolic tangent function.

An LSTM model, operating as a black-box method, has the potential to exhibit overfitting issues, diminishing its effectiveness when applied to new, untrained data. To gauge the extent of overfitting, data is typically divided into two subsets in the realm of machine learning: the training set and the validation set. The training set is utilized for model development, while the validation set remains untouched during the training phase and serves to evaluate the model's predictive performance.

Traditionally, practitioners have favored training their models on large datasets with numerous observations, a practice grounded in the law of large numbers. However, I argue that this conventional approach may not be well-suited for forecasting financial variables. In essence, I posit that a model trained on the daily prices of an asset spanning the years 1940 to 2022 may not perform as effectively as a model trained solely on the daily prices between 2020 and 2022.

One potential reason for this discrepancy lies in the fact that a long-range training set encompasses both downward and upward market trends. Such training data may not yield accurate predictions when applied to data sampled during a trend in a single direction. Consequently, utilizing more recent data points as the training set may lead to superior predictive performance.

In this paper, I propose the use of rolling training sets for predicting the next observation. In this scenario, the 5,300th, 5,301st, 5,302nd, 5,303rd, and 5,304th observations serve as the training set to forecast the 5,305th observation when the training window is set to 5. This approach ensures that every observation is predicted based on the most recent price action, rather than relying on price action from hundreds of days ago.

To test the effectiveness of this new approach, I used 2 training set rules:

*Approach 1 (MECE):* The entire dataset is divided into two mutually exclusive and collectively exhaustive sets, namely a training set and a test set. I used the first 5,282 observations as the training set and the remaining 300 observations are used as the test set. In this case, I forecasted all observations among the last 300 observations based on a single model developed by using the first 5,282 observations.

*Approach 2 (Rolling):* For every observation among the last 300 observations, the prior 5, 10, 20, and 50 observations are used as the training set. In this approach, a model is trained to forecast the next observation for each training window.

In total, I trained 2,408 LSTM models with configurations above to forecast the last 300 observations in the sample. As a preprocessing step, I performed a transformation on the daily mid-prices by subtracting 100 from each value and then scaling the results by a factor of 1/100, resulting in the formula x/100-1. This scaling operation has the effect of confining all observations within the range of -1 to 1, consistent with the range of tanh function used in LSTM models. It's important to note that this scaling choice is based on the assumption that the daily mid-prices will consistently remain below 100 Turkish Lira (TRY). The selection of this threshold is based on the observation that all data points within the fixed-range training set are significantly lower than the chosen threshold. Consequently, I have intentionally refrained from constraining the model to only produce forecasts that surpass the maximum value observed in the training set.

I made a deliberate decision to avoid using the conventional min-max scaling method. This choice was driven by the understanding that min-max scaling assumes prior knowledge of the range of daily mid-prices in the validation set. However, in this context, the range of these mid-prices is considered unknown since they are the very values we aim to forecast.

## 3. Results

### 3.1. Historical Premiums

Figure 3 shows that KRDMA was traded at a premium relative to KRDMB for only 1,171 days out of a total of 5,582 days. In 2020, the premium of KRDMA over KRDMB was the strongest when KRDMA predominantly traded at a premium for most of that year. Conversely, KRDMD consistently saw substantial discounts relative to both KRDMA and KRDMB. Over a span of 4,360 days, KRDMA was traded at a premium. Similarly, KRDMB was traded at a premium status for 4,318 days. Figure 1 also demonstrates a substantial decrease in premiums paid for KRDMA and KRDMB over KRDMD, starting from 2018. A reversion occurred starting in 2021 for KRDMD discounts. Since 2021, both KRDMA and KRDMB have consistently been traded at a discount relative to KRDMD.

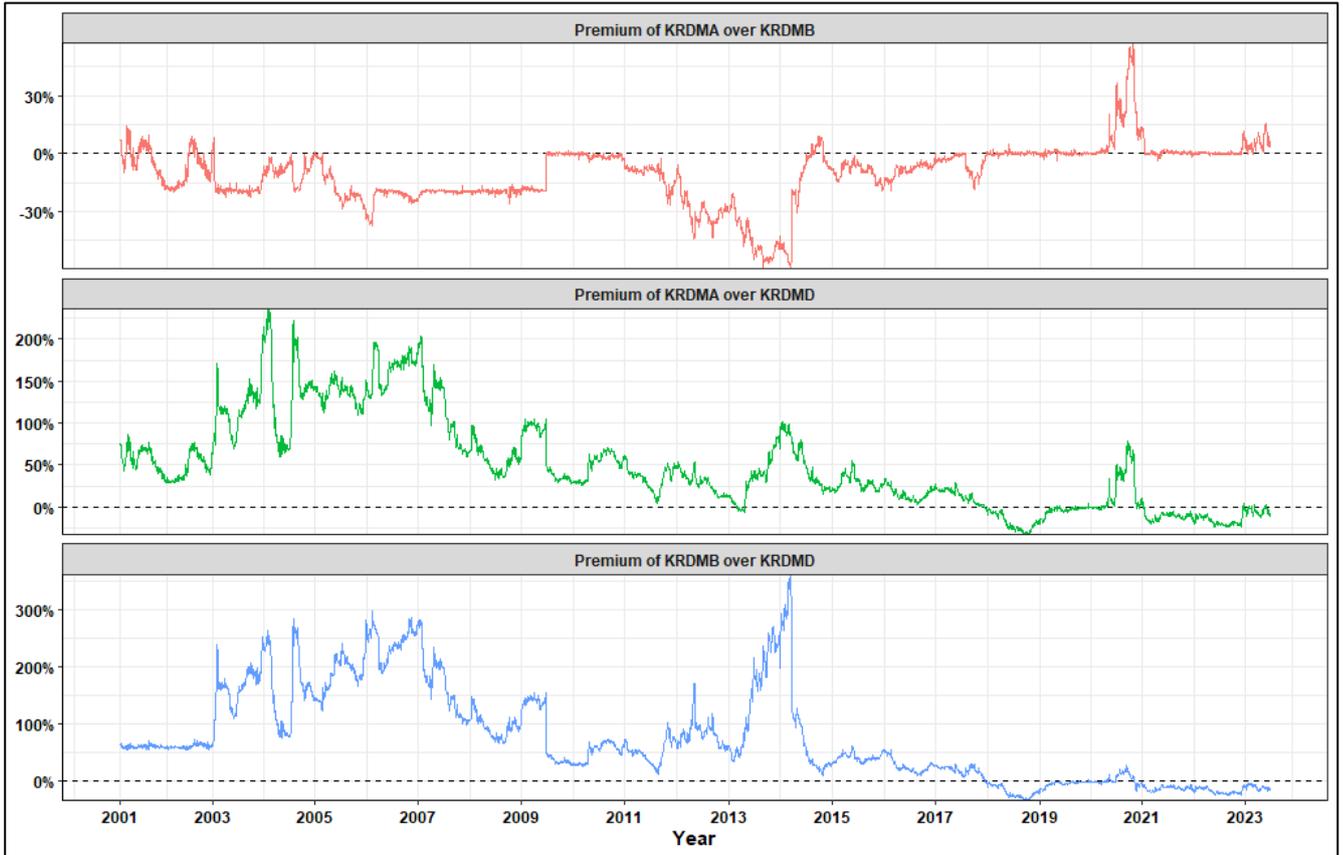

**Figure 3: Historical Premiums**

Summary statistics of these premiums are given in Table 1. It clearly shows that substantial premiums paid for KRDMA and KRDMB over KRDMD between 2001 and 2023. At their heights these premiums reached 235.87% and 361.21% respectively for KRMDA and KRDMB.

**Table 1.** Summary Statistics of Premiums

|  | *KRDMA over KRDMB* | *KRDMA over KRDMD* | *KRDMB over KRDMD* |
|---|---|---|---|
| **Minimum** | -59.70% | -33.08% | -33.21% |
| **1st Quartile** | -19.10% | 5.99% | 12.98% |
| **Median** | -7.71% | 35.47% | 57.82% |
| **Mean** | -10.30% | 49.44% | 75.73% |
| **3rd Quartile** | 0.00% | 73.51% | 128.13% |
| **Maximum** | 57.06% | 235.87% | 361.21% |

**3.2. Wavelet Coherence Analysis**

In this section, I present the dynamic relationship between daily returns (calculated as percentage change in daily mid-prices) of Kardemir stocks by using the wavelet coherence technique explained above.

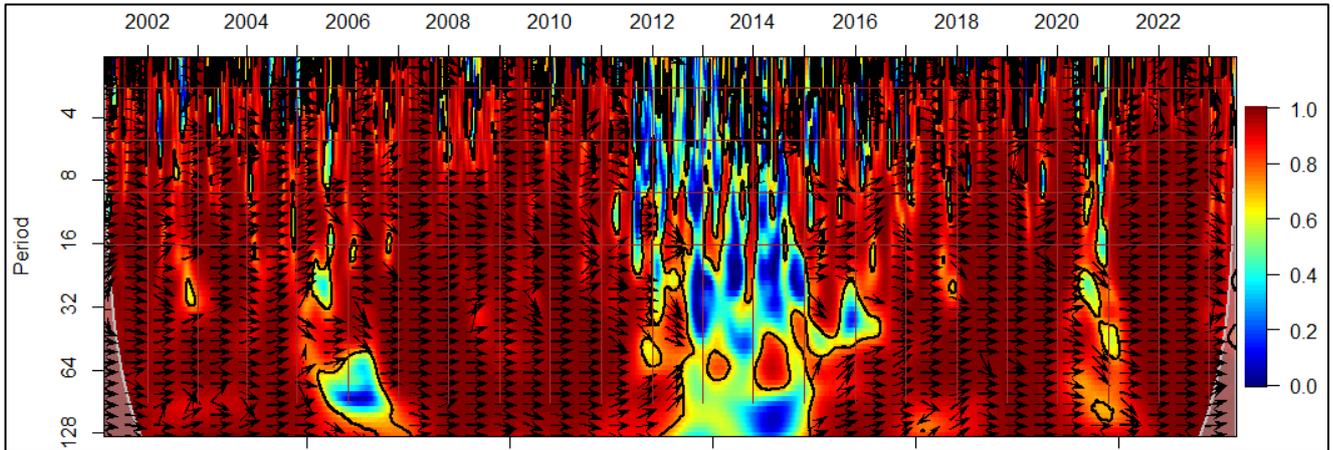

**Figure 4: Wavelet Coherence between KRDMA and KRDMB Daily Returns**

As depicted in Figure 4, the coherence between the daily returns of KRDMA and KRDMB remained consistently strong throughout the analyzed period, with only a few exceptions. The first notable deviation occurred in 2006, spanning periods 64 to 128, during which there was a clear lack of coherence between the daily return patterns of the two stocks. The second significant divergence surfaced in 2012 and persisted for over two years. During this period, the daily returns of KRDMA and KRDMB exhibited noticeable discrepancies. It's worth highlighting that over this time frame, the premium of KRDMA over KRDMB reached its lowest point, showing a substantial decline of -59.64%. Importantly, the figure also emphasizes the absence of a substantial cause-and-effect relationship between these two series. Instead, their temporal progression displayed synchronized movements, without any prominent identifiable temporal precedence or lag.

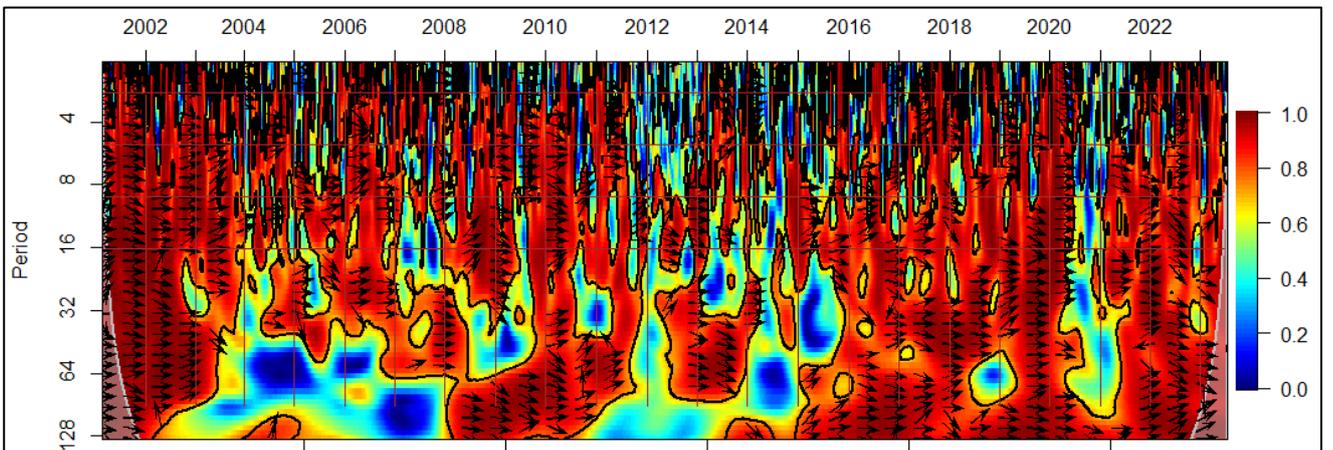

**Figure 5: Wavelet Coherence between KRDMA and KRDMD Daily Returns**

Figure 5 illustrates the results of the wavelet coherence analysis applied to the relationship between KRDMA and KRDMD. Across most of the time span, from January 2001 to July 2023, the daily returns of these two stocks displayed significant synchronization. It's worth noting that during this period, the coherence between the two stocks was notably strong and consistent, contributing to their aligned behavior.

Valuable insights can be derived from the coherence patterns at various time scales. Specifically, between 2003 and 2008, as well as intermittently in the first quarter of 2012 and throughout 2014, the coherence associated with longer cycles, particularly those spanning from 64 to 128 days, was relatively weak when compared to the coherence observed within shorter cycles, ranging from 8 to 16 days. This observation underscores temporal variations in the degree of synchronization across different scales, highlighting periods of heightened and diminished shared behavior.

Moreover, a clear absence of coherence becomes evident in the latter part of 2020 across various time cycles. During this specific period, the synchronization between the two stocks was notably absent. What's particularly noteworthy is that this timeframe coincided with a substantial premium of over 75% attributed to KRDMA over KRDMD. This convergence of factors highlights the potential interplay between coherence patterns and premium fluctuations, implying intricate dynamics at play in the relationship between these stock returns during this period.

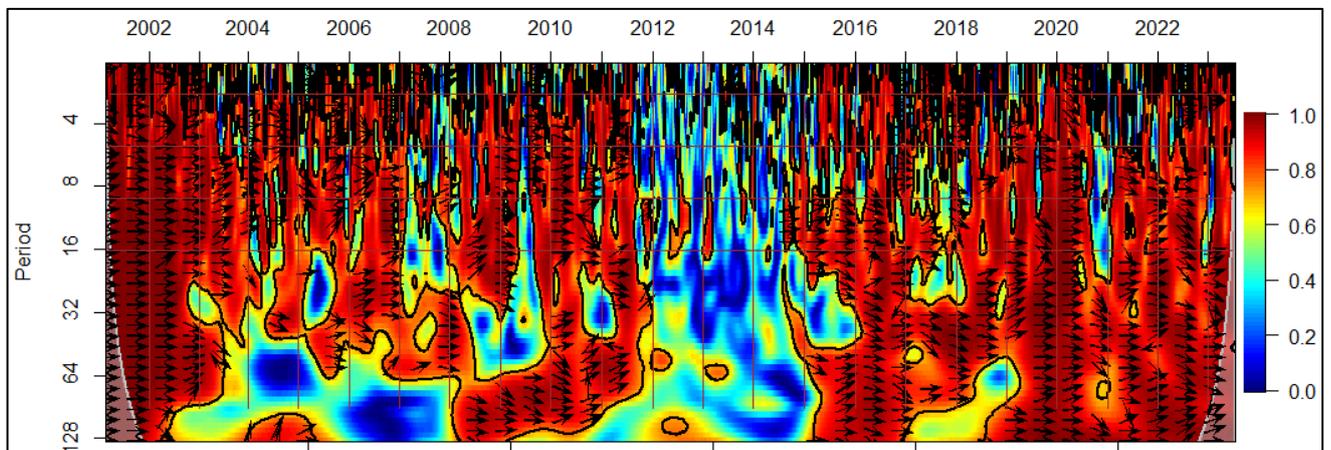

**Figure 6: Wavelet Coherence between KRDMB and KRDMD Daily Returns**

The final wavelet coherence plot, depicted as Figure 6, reveals the weakest coherence observed among Kardemir stock returns, specifically between KRDMB and KRDMD. Notably, the coherence between the daily returns of KRDMB and KRDMD was particularly weak, primarily spanning the years from 2012 to 2015. This period coincided with a time when the premium paid for KRDMB over KRDMD reached its peak, skyrocketing to an unprecedented level of 361.21%.

Across the broader time frame spanning from January 2001 to 2012, a robust coherence pattern was evident among the daily returns of the three Kardemir stocks across various time cycles. This robust coherence paradigm underwent a transformation, transitioning to a less robust coherence configuration from 2012 to 2015, only to reemerge in 2016. Another episode of coherence weakness emerged towards the latter part of 2020, encompassing all Kardemir stocks within shorter cycles. Typically, the highest coherence was observed between KRDMA and KRDMB.

Furthermore, the wavelet plots emphasize the absence of a clearly discernible leading or lagging relationship between various Kardemir stocks. Notably, instances of significant divergence in the daily mid-prices of Kardemir stocks coincided with the absence of coherence, particularly within longer cycles. This observation suggests a potential complex interplay between coherence dynamics and price disparities in the long-term context**.**

### 3.3. Long Short-Term Memory (LSTM) Models

In this section, I present the predictive performance results obtained from a total of 2,408 LSTM models. These models were trained with the specific goal of forecasting the daily mid-prices of the latest 300 observations. To evaluate the effectiveness of dual-class stocks as predictors for each other, we employed two sets of models.

In the first set of models, the prediction models did not incorporate the historical price movements of dual stocks as lagged variables when forecasting future mid-prices. In contrast, the second set of models was designed to include lagged dual-class stock mid-prices as factors for predicting future mid-prices. For these lagged variables, I considered two options: using 4 lags and 9 lags of daily mid-prices. All these models were trained following one of the five training-set rules outlined in Section 0 above. To evaluate and compare their predictive performance, I employed three key metrics: root mean squared error (RMSE), mean absolute error (MAE), and mean absolute percentage error (MAPE).

**Table 2. KRDMA Predictive Performance Results**

|  | Lag = 4 | | Lag = 9 | |
| --- | --- | --- | --- | --- |
| *Training Window = 5* | *Dual-Stock = No* | *Dual-Stock = Yes* | *Dual-Stock = No* | *Dual-Stock = Yes* |
| **RMSE** | 0.7433 | 0.8106 | 0.8006 | 0.8080 |
| **MAE** | 0.5077 | 0.5466 | 0.5578 | 0.5575 |
| **MAPE** | 3.387 | 3.6617 | 3.7375 | 3.7516 |
| *Training Window = 10* | *Dual-Stock = No* | *Dual-Stock = Yes* | *Dual-Stock = No* | *Dual-Stock = Yes* |
| **RMSE** | 0.9027 | 0.9698 | 1.0020 | 1.1226 |
| **MAE** | 0.6230 | 0.6745 | 0.7044 | 0.7667 |
| **MAPE** | 4.2066 | 4.5580 | 4.8132 | 5.2047 |
| *Training Window = 20* | *Dual-Stock = No* | *Dual-Stock = Yes* | *Dual-Stock = No* | *Dual-Stock = Yes* |
| **RMSE** | 1.1606 | 1.1915 | 1.4265 | 1.4671 |
| **MAE** | 0.8411 | 0.7863 | 0.9940 | 1.0660 |
| **MAPE** | 5.7301 | 5.3693 | 6.8049 | 7.3317 |
| *Training Window = 50* | *Dual-Stock = No* | *Dual-Stock = Yes* | *Dual-Stock = No* | *Dual-Stock = Yes* |
| **RMSE** | 1.4709 | 1.4571 | 1.6515 | 1.8731 |
| **MAE** | 1.1134 | 1.0881 | 1.2367 | 1.3784 |
| **MAPE** | 8.0633 | 7.7143 | 8.9525 | 9.8199 |
| *MECE* | *Dual-Stock = No* | *Dual-Stock = Yes* | *Dual-Stock = No* | *Dual-Stock = Yes* |
| **RMSE** | 0.8173 | 0.8510 | 1.1949 | 1.0517 |
| **MAE** | 0.5873 | 0.6132 | 0.9246 | 0.7664 |
| **MAPE** | 3.9752 | 4.1735 | 6.3747 | 5.2103 |

**Table 3. KRDMB Predictive Performance Results**

|  | Lag = 4 | | Lag = 9 | |
| --- | --- | --- | --- | --- |
| *Training Window = 5* | *Dual-Stock = No* | *Dual-Stock = Yes* | *Dual-Stock = No* | *Dual-Stock = Yes* |
| **RMSE** | 0.6834 | 0.7567 | 0.7440 | 0.7757 |
| **MAE** | 0.4601 | 0.4995 | 0.5047 | 0.5251 |
| **MAPE** | 3.2067 | 3.4693 | 3.5233 | 3.6607 |
| *Training Window = 10* | *Dual-Stock = No* | *Dual-Stock = Yes* | *Dual-Stock = No* | *Dual-Stock = Yes* |
| **RMSE** | 0.8587 | 0.8917 | 1.0084 | 1.0299 |
| **MAE** | 0.5756 | 0.6030 | 0.6862 | 0.7005 |
| **MAPE** | 4.0155 | 4.2513 | 4.8465 | 4.9389 |
| *Training Window = 20* | *Dual-Stock = No* | *Dual-Stock = Yes* | *Dual-Stock = No* | *Dual-Stock = Yes* |
| **RMSE** | 1.1757 | 1.1435 | 1.3400 | 1.3932 |
| **MAE** | 0.7796 | 0.7850 | 0.8755 | 0.9492 |
| **MAPE** | 5.4603 | 5.5653 | 6.2216 | 6.6879 |
| *Training Window = 50* | *Dual-Stock = No* | *Dual-Stock = Yes* | *Dual-Stock = No* | *Dual-Stock = Yes* |
| **RMSE** | 1.4014 | 1.4328 | 1.6128 | 1.6676 |
| **MAE** | 1.0322 | 1.0424 | 1.2265 | 1.2750 |
| **MAPE** | 7.6067 | 7.6053 | 9.0592 | 9.5743 |
| *MECE* | *Dual-Stock = No* | *Dual-Stock = Yes* | *Dual-Stock = No* | *Dual-Stock = Yes* |
| **RMSE** | 0.7612 | 0.7824 | 0.9884 | 0.9943 |
| **MAE** | 0.5407 | 0.5490 | 0.7306 | 0.7220 |
| **MAPE** | 3.8301 | 3.9241 | 5.2094 | 5.0590 |

**Table 4. KRDMD Predictive Performance Results**

|  | Lag = 4 | | Lag = 9 | |
| --- | --- | --- | --- | --- |
| *Training Window = 5* | *Dual-Stock = No* | *Dual-Stock = Yes* | *Dual-Stock = No* | *Dual-Stock = Yes* |
| **RMSE** | 0.7910 | 0.8321 | 0.8596 | 0.8564 |
| **MAE** | 0.5608 | 0.5811 | 0.6050 | 0.6117 |
| **MAPE** | 3.4491 | 3.5723 | 3.7123 | 3.7537 |
| *Training Window = 10* | *Dual-Stock = No* | *Dual-Stock = Yes* | *Dual-Stock = No* | *Dual-Stock = Yes* |
| **RMSE** | 0.9408 | 1.0293 | 1.1151 | 1.1721 |
| **MAE** | 0.6584 | 0.7155 | 0.7678 | 0.8209 |
| **MAPE** | 4.0577 | 4.4043 | 4.7642 | 5.0716 |
| *Training Window = 20* | *Dual-Stock = No* | *Dual-Stock = Yes* | *Dual-Stock = No* | *Dual-Stock = Yes* |
| **RMSE** | 1.3288 | 1.4243 | 1.4513 | 1.4485 |
| **MAE** | 0.9194 | 1.0142 | 1.0385 | 1.0465 |
| **MAPE** | 5.6409 | 6.3443 | 6.4796 | 6.5548 |
| *Training Window = 50* | *Dual-Stock = No* | *Dual-Stock = Yes* | *Dual-Stock = No* | *Dual-Stock = Yes* |
| **RMSE** | 1.4489 | 1.6598 | 1.6948 | 1.9712 |
| **MAE** | 1.1061 | 1.2782 | 1.2880 | 1.5104 |
| **MAPE** | 7.1237 | 8.1918 | 8.1252 | 9.7193 |
| *MECE* | *Dual-Stock = No* | *Dual-Stock = Yes* | *Dual-Stock = No* | *Dual-Stock = Yes* |
| **RMSE** | 0.8730 | 0.9280 | 1.2142 | 1.1088 |
| **MAE** | 0.6322 | 0.6894 | 0.9765 | 0.8100 |
| **MAPE** | 3.9094 | 4.2723 | 6.3353 | 5.0450 |

Tables 2–4 illustrate that including lagged daily mid-prices of dual-class stocks as predictors, alongside the lagged daily mid-prices of individual stocks, did not result in an improvement of predictive performance. In fact, in most cases, I observed a modest decline in predictive accuracy when incorporating dual stocks as predictors. In certain instances, any improvements were minimal at best. It's worth highlighting that the validation set, comprising the last 300 data points, coincided with a period marked by strong coherence between the daily returns of KRDMA, KRDMB, and KRDMD. Despite this robust coherence, the inclusion of dual stocks did not yield any noticeable additional insights into future stock prices.

Furthermore, the tables also demonstrate that the model utilizing a 5-day training window consistently outperformed other models in all predictive performance metrics. This finding aligns with my thesis opposing the use of extended training periods in financial data forecasting and is in line with prior research findings regarding investor and managerial myopia [10], [4], [7], and [27].

## 4. Conclusions

In this study, I delve into a case of dual-class stock structure on Borsa Istanbul, marked by a distinctive characteristic: prolonged disparities among three stocks, namely KRDMA, KRDMB, and KRDMD. These disparities reached staggering heights, with differentials soaring as high as 361.21%. To put this into perspective, consider that the most significant divergence observed between GOOG and GOOGL on NASDAQ has been approximately 5% since 2005. This stark contrast underscores that arbitrage opportunities between KRDMA, KRDMB, and KRDMD may present greater profit potential, albeit potentially slower returns compared to other publicly traded companies with dual-class stock structures.

Moreover, this study also sheds light on the fact that even when there exists a robust coherence between dual-class stock prices, these prices may not necessarily serve as reliable predictors for future price movements. Lastly, I offer substantial empirical evidence supporting the practice of favoring shorter training periods over extended ones, contrary to the common practice of employing larger training sets with numerous observations and lags. These findings hold valuable implications for practitioners seeking to enhance the predictive performance of machine learning models in financial applications.